# The bi-pyramidal nature, the Lucas series in the genetic code and their relation to aminoacyl-tRNA synthetases

Chi Ming Yang, Ph.D.

Physical Organic Chemistry and Chemical Biology, Nankai University, Tian Jin 30007, China.   E-mail: yangchm@nankai.edu.cn   Fax: 011 86 22 2350 3863

*(July, 2003)*

**Abstract**

It has been unclear what principle governs the selection of the 20 canonical amino acids in the genetic code. Based on a previous study of the 28-gonal and rotational symmetric arrangement of the 20 amino acids in the genetic code, new analyses of the organization of the genetic code system together with their relation to the two classes of aminoacyl-tRNA synthetases are reported in this work. A closer inspection revealed how the enzymes and the 20 amino acids of the genetic code are intertwined on a polyhedron model. Complimentarily and cooperative symmetry between class I and class II aminoacyl-tRNA synthetases displayed by a 28-gon model are discussed, and we found that the two previously suggested evolutionary axes of the genetic code overlap the three two-fold symmetry axes within the two classes of aminoacyl-tRNA synthetases. Moreover, it is identified that the amino-acid side-chain carbon-atom numbers (1, 3, 4 and 7) in the overwhelming majority of the amino acids recognized by each of the two classes of aminoacyl-tRNA synthetases meet a mathematical relationship, the Lucas series.

1. **Introduction**

   Since Weber and Miller (1981) described reasons for the occurrence of the twenty coded protein amino acids, the fundamental properties of the 20 canonical amino acids have since inspired tremendous amount of excellent research and speculation (Davydov, 1998; Dufton, 1997; Rakočević and Jokic, 1996; Rakočević, 1998; Shcherbak, 1989).  The genetic code was more recently shown to be determined by Golden mean through the unity of the binary-code tree and the Farey tree in Rakočević's work (1998), and it was identified that atom number balance in amino acids are directed by Golden mean route, also directed by the double-triple system of amino acids, as well as by two classes of aminoacyl-t-RNA synthetases (AARS's).

   As is well known, the first two bases in the tri-nucleotide codons are recognized to be the most specific fragments of the codons. Therefore, they are instrumental for analysis aimed at elucidation of the coding principle underlying the genetic code system.  The 64 codons can be divided into 16 groups of genetic code doublets, such as UUN, UCN, UGN, UAN…., N = U, C, G and A. We recently explored a structural regularity within the four RNA nucleobases U, C, G and A by invoking the nature of covalent bonding, the $sp^2$-hybrid nitrogen-atom numbers in the nucleobase molecules, to investigate the symmetry inherent in the genetic code (Yang, 2003a). By advancing and modifying some early graphic and geometric approaches to understanding the genetic code (Karasev and Stefanov, 2001; Yang, 2003b), solid geometric analysis of the 16 groups of genetic code doublets revealed that the rotational symmetries inherent in the distribution of both the number of the amino acids and their side-chain carbon-atom contents in the genetic code follow a quasi-28-gon (icosikaioctago) model with two presumed symmetry axes (Yang, 2003b; 2003c). The newly identified polyhedral symmetric nature as a 28-gonal pyramid, or a double pyramidal nature (Yang, 2003b), of the genetic code, echoes the similar, but different structure of the simplest viruses, *i.e.*, spherical viruses, which is of icosahedral symmetry (Casper and Klug, 1962; Racaniello, 1996).

The overall symmetric feature of the code described by our spherical and polyhedral model together with two presumed evolutionary axes is in an excellent agreement with the recent Trifonov's proposal that Ala could be the first codonic amino acid (Trifonov and Bettecken, 1997; Trifonov, 2000). From the polyhedron symmetric model, we can also identify that the symmetric distribution of the 20 amino acids around the 28-gon excellently fits the Rumer's regularity (Shcherbak, 1989). Both the amino acids in the Group IV (degeneracy 4) and the rest of the amino acids in the Quasi-group III-II-I are each occupying half of the surface on the 28-gonal model (Yang, 2003), see Figure 1.

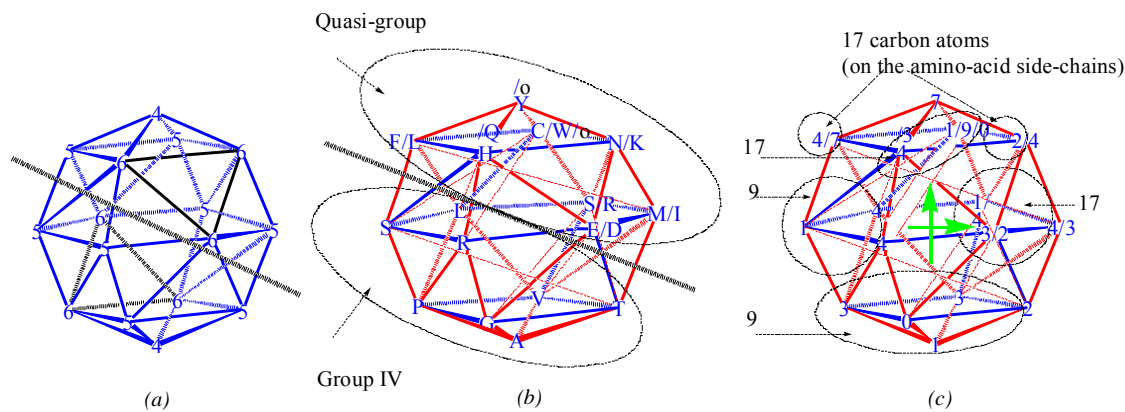

**Figure 1**. The icosikaioctagonal or 28-gonal relation within the 20 amino acids in the genetic code. *(a)* An icosikaioctagon or a 28-gon. The numbers at vertices indicate the edges at the vertices. *(b)* The distribution of the 20 amino acids is symmetric following a 28-gon model. *(c)* The side-chain carbon-atom numbers (SCCAN) within the amino acid molecules at each of the 16 genetic code doublets. Two presumed symmetry axes are indicated by arrows, "⇑" and "⇒". Block lines (both red and blue) are edges on the polyhedron; red lines (both block and dotted) are for neighboring code-doublet connection.

The establishment of the genetic code essentially requires 20 enzymes aminoacyl-tRNA synthetases (AARS's), which catalyze aminoacylation of tRNAs by joining an amino acid to its cognate tRNA to establish the specificity of protein synthesis (de Pouplana and Schimmel, 2001; Hartman, 1995; Hartlein and Cusack, 1995; Cavalcanti et al., 2000; Di Giulio, 1992; Woese et al., 2000). In the present work we extend the study to the cooperative symmetry within the AARS's in response to the polyhedral and rotational symmetry of the genetic code system, and seek an explanation for the possibly intertwined symmetry between the genetic code and AARS's. Finally, we reach a conclusion by identifying the Lucas series, from the carbon-atom numbers in amino acid side-chains, and propose that the mathematical principle as the Lucas series may have been underlying the natural selection of the amino-acid contents in the genetic code.

## 2 A polyhedral symmetric model displaying how the genetic code may be intertwined with AARS's

AARS's possess very high specificity in selecting their cognate amino acid and tRNA substrates. Presently, AARS's have been classified into two distinct while structurally unrelated classes (I and II), each class encompassing 10 amino acids (Woese, 2000), see Table 1. A well-known assumption is that the tRNA-charging function of the AARS's evolved at least twice. Perhaps the two classes reflect a dichotomous origin of protein translation processes via two different primitive processes (Woese, 2000; Nagel and Doolittle, 1991). Therefore, 20 AARS's, which are responsible for establishing the genetic code, are the potential markers of genetic code development (de Pouplana and Schimmel, 2001). Consequently, it has been suggested that the evolution of the genetic code and evolution of the AARS enzymes are intertwined (Woese, 2000), and accordingly, the evolution of the AARS's are instrumental in understanding the selective pressures maintaining the genetic code.

The coexistence of the two distinct classes of AARS's is one of the most striking features of the AARS's (Arnez and Moras, 1997; Burbaum and Schimmel, 1991; Cusack, 1997; Eriani et al., 1990). However, despite a clear existence of correlation between the genetic code and the evolutionary patterns of the AARS's, the biological importance of this fact is not known. According to this reasoning, an investigation into the internal relation of AARS's with respect to the symmetric relation within the amino acids is likely to provide insight into the conservation and evolution of the genetic code.

**Table 1.** Two classes of aminoacyl-tRNA synthetases corresponding to two classes of amino acids.[a]

| AARS Class / Presumed Evolutionary Stage | Class I | Class II |
|---|---|---|
| Stage 3 | TyrRS<br>TrpRS<br>CysRS<br>GlnRS<br>LysRS-I | PheRS<br>AsnRS<br>HisRS<br>LysRS-II |
| Stage 2 | LeuRS<br>ArgRS<br>MetRS<br>IleRS<br>GluRS | AspRS<br>SerRS |
| Stage 1 | ValRS | ProRS<br>ThrRS<br>GlyRS<br>AlaRS |

[a] Here, ArgRS represents arginyl-tRNA synthetase, and so forth.
Two types of lysyl-tRNA synthetases (class I and class II) are labeled accordingly.

Given the distribution of the 20 amino acids is symmetric following a 28-gon model (Yang, 2003b), we now start from examining how the polyhedral symmetric relation of the 20 amino acids within the genetic code and the AARS's are intertwined on the polyhedral symmetric model (Figure 2).

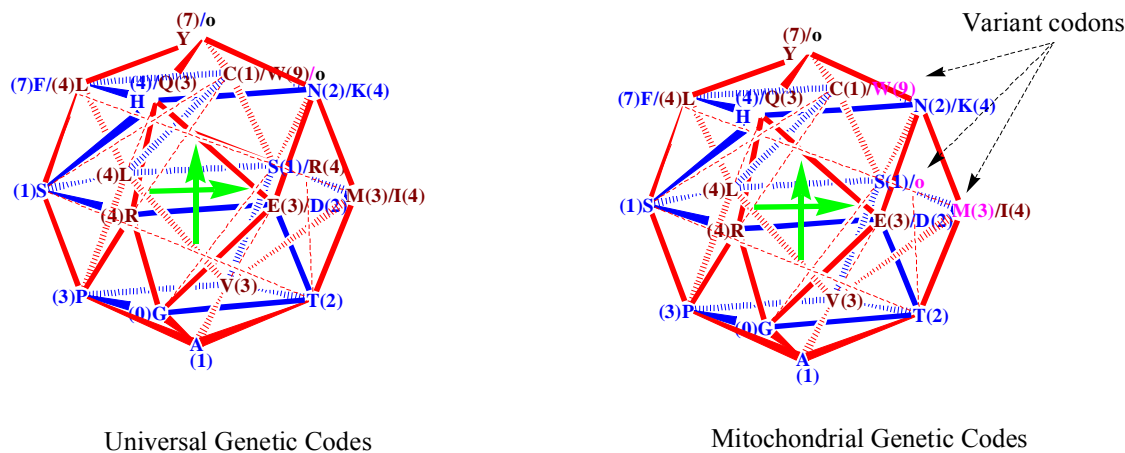

Red brown: Calss I aminoacyl-tRNA synthetase
Blue: Calss II aminoacyl-tRNA synthetase

Differences between the Universal and Mitochondrial Genetic Codes

| Codon | Universal code | Human mitochondrial code |
|---|---|---|
| UGA | Stop | Trp |
| AGA | Arg | Stop |
| AGG | Arg | Stop |
| AUA | Ile | Met |

**Figure 2.** Amino acids, recognized by two classes of aminoacyl-tRNA synthetases, corresponding to the universal genetic code and mitochondria genetic code on an icosikaioctagon or a 28-gon model. The numbers at vertices indicate the edges at the vertices. Shown on the side of amino-acid letters are the side-chain carbon-atom numbers (SCCAN) within the amino acid molecules. Two presumed symmetry axes are indicated by arrows, "⇑" and "⇒". Block lines (both red and blue) are edges on the polyhedron; red lines (both block and dotted) are for neighboring code-doublet connection.

In Figure 2, the intertwining of the genetic code with AARS's can be clearly displayed on the 28-gon. In an effort to explore the internal symmetric relation of two classes of AARS's and the 20 amino acids, we divide the whole system into three presumed stages for the genetic code being intertwined with AARS's on the polyhedral symmetric mode. As is evident from Figure 3, there are three two-fold symmetry axes within the two classes of AARS's, *i.e.*, one two-fold symmetry axis within every presumed evolutionary stage.

The display helps understand that the close evolutionary relation between the (class I, II) AARS's is mirrored in the obvious coding relationship between the corresponding amino acids and the relationship between their codons. Although the degeneracies of all the 20 amino acids vary, exactly half of the 16 genetic code doublets positions are recognized by one of the two classes of AARS's, Figure 3.

It has been pointed out that the central role played by the AARSs in translation may suggest that their evolution and that of the genetic code are somehow intertwined. Hence, the finding in the present work may provide a clue to answer the question of whether the AARS's in their evolution have contributed to the code's present structure, or are the codon assignments simply reflections of AARS's evolutionary wanderings (Woese, 2000)?

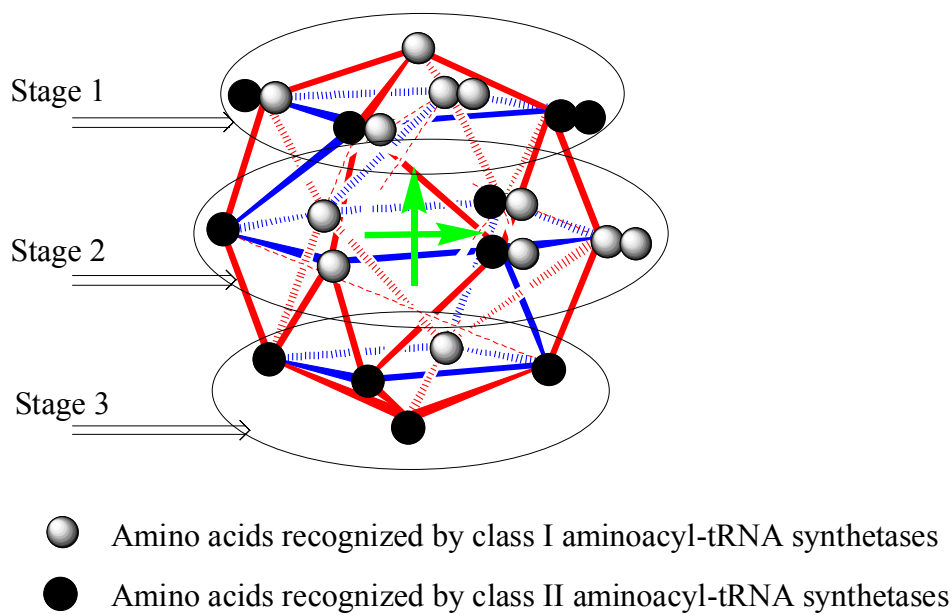

○ Amino acids recognized by class I aminoacyl-tRNA synthetases

● Amino acids recognized by class II aminoacyl-tRNA synthetases

**Figure 3.** Distribution of the two classes of amino acids are also symmetric with a two-fold symmetric axis at each of the three presumed stages for the genetic code, which is intertwined with aminoacyl-tRNA synthetases on the polyhedral symmetric model. (o, Amino acids recognized by class I aminoacyl-tRNA synthetases; ●, Amino acids recognized by class II aminoacyl-tRNA synthetases. The special case is lysine, which in some organisms is charged by a class I enzyme.)

We now analyze amino acid side-chain carbon-atom number balance based on the polyhedral symmetry of the genetic code with respect to the two classes of AARS's. The results are shown in Figure 4.

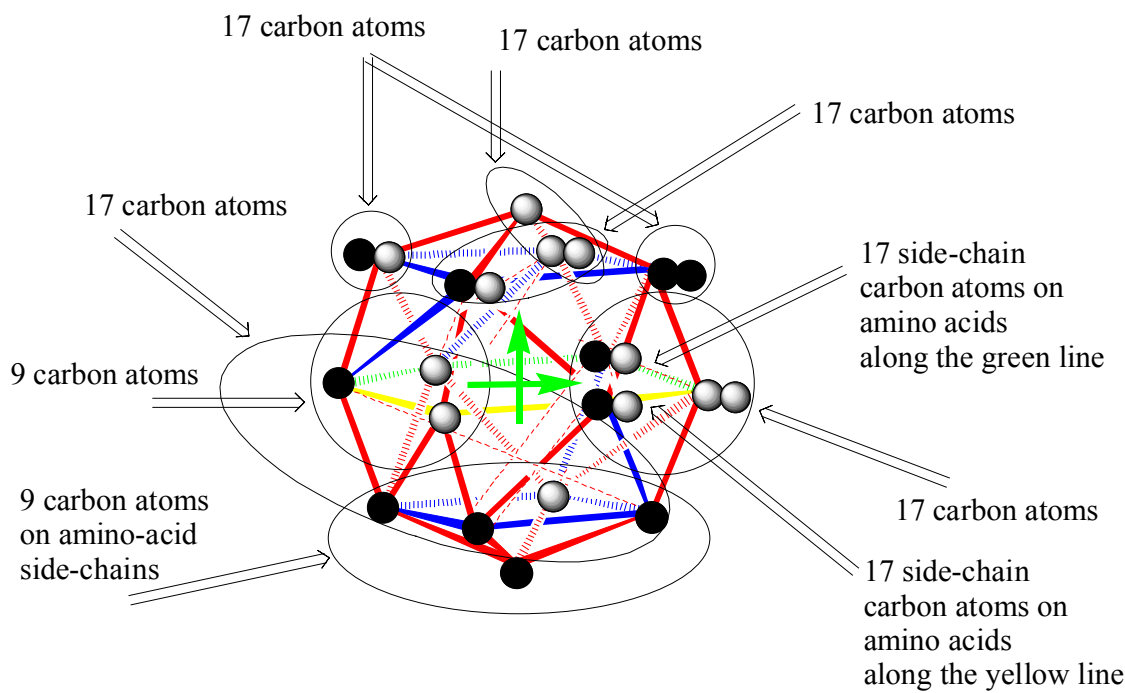

**Figure 4.** Amino-acid side-chain carbon-atom number balance summarized on the polyhedral symmetric model and their relation to the two classes of aminoacyl-tRNA synthetases. We can see the unique number "17" in several groups of amino acids. (o, Amino acids recognized by class I aminoacyl-tRNA synthetases; •, Amino acids recognized by class II aminoacyl-tRNA synthetases. The special case is lysine, which in some organisms is charged by a class I enzyme.)

Depicted in Figure 3 and 4, after we deploy the 20 amino acids on the polyhedral model in response to class I and II AARS's, the two presumed evolutionary axes within the genetic code are more evident. First, the 20 amino acids can also be grouped into 3 evolutionary stages, *i.e.*, 1) A, P, G, V and T; 2) S, L, R, E, D, M and I; 3) F, L, H, Q, N, K, C, W and Y. Second, at each stage on the polyhedral model, there is one C-2 symmetric plane with respect to the two classes of AARS's, see Figure 3. Also of note is the fact that the total numbers of amino-acid side-chain carbon atoms within each stage are 9, 9+17, 17+17+7, respectively, Figure 4.

From this step-by-step graphic examination of the AARS's and amino-acid contents within the genetic code, it is clear that the both the symmetric and possible evolutionary relation between the genetic code and AARS's are closely coiled with each other. In addition to the amino-acid side-chain carbon atom number balance at "17" indicated at several positions at stages 2 and 3 in Figure 4, amino-acid side-chain carbon-atom number balance at "17" also occurs at

*F(7); N(2)/K(4); H(4)* ≡ Q(3); C(1)/W(9); L(4)

On the left of the 'equation' are the amino acids recognized by class II AARS's, on the right are the amino acids by class I. We will discuss this stranger but interesting number "17" in *Section 3*.

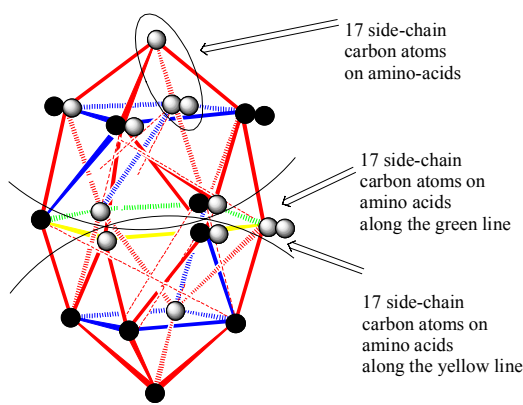

**Figure 5**. The bi-pyramidal nature of the genetic code, displayed by the amino-acid side-chain carbon-atom number balance on the polyhedral (icosikaioctagon or 28-gon) model when the 20 amino acids in the genetic code are viewed as a whole.

Provided that the 20 canonical amino acids in the genetic code can be deployed on the surface of a 28-gon model, the above atomic content analysis also revealed a bi-pyramidal nature of the genetic code, shown in Figure 5a. Based on the relative atomic compositions in amino acids, see Figure 5b, vertically, from A⇒P⇒S⇒F/L⇒Y, the total number of the amino acid side-chain carbon-atoms equals to 23; from A⇒T⇒M/I⇒N/K⇒Y, the total number of the amino acid side-chain carbon-atoms equals to 23; by comparison, horizontally, from S⇒R⇒E/D⇒M/I, the total number of the amino acid side-chain carbon-atoms equals to 17; from S⇒L⇒S/R⇒M/I, the total number of the amino acid side-chain carbon-atoms equals to 17. A ratio of 15 [(P + S + F/L) or (T + M/I + N/K)] to 9 [(R + E/D) or (L + S/R)] equals to 1.6666, which is close the Golden mean 1.6180339…

This finding helps explain why the genetic code is one of the most highly conserved characters in living organisms.

### 3. A Fibonacci-Lucas relationship within the canonical amino acids

We now explore the possible existence of a mathematical relationship within the canonical amino acids.

### 3.1 The Fibonacci series

Certain mathematical concepts of similarity and proportion hold one of the keys to understanding processes of growth in the natural world. For example, the Fibonacci series: 1, 1, 2, 3, 5, 8, …, is well known to lie at the heart of plant growth and living organisms (Kappraff and Marzec, 1983). The Fibonacci sequence is generated by adding the previous two numbers in the list together to form the next and so on and so on...

$$0, 1, 1, 2, 3, 5, 8, 13, 21, 34, 55, 89, 144, ...$$

The basic Fibonacci relationship is in eq. 1:

$$F_{i+2} = F_{i+1} + F_i \qquad (1)$$

In 1202, Leonardo Fibonacci (1170-1240) in Italy discovered this simple numerical series that is the foundation for a remarkable mathematical relationship behind Phi or Φ. Starting with 0 and 1, each new number in the series is simply the sum of the two before it. Divide any number in the Fibonacci sequence by the one before it, for example 55/34, or 21/13, and the answer quickly converge on the Golden mean Phi or Φ, 1.618033988749895 . . .. After the 40th number in the series, the ratio is accurate to 15 decimal places.

**3.2 The Lucas series**

A French mathematician, Edouard Lucas (1842-1891), who gave the above series of numbers 0, 1, 1, 2, 3, 5, 8, 13, .. the name *the Fibonacci Numbers*, found a similar series occurs often when investigating Fibonacci number patterns:

$$2, 1, 3, 4, 7, 11, 18, 29, 47, 76, 123, ...$$

Here, the Fibonacci rule of adding the latest two to get the next is kept, but it starts from 2 and 1 (in this order) instead of 0 and 1 for the ordinary Fibonacci numbers.

This series, called the Lucas Numbers, is defined as follows: where we write its members as $L_i$, for Lucas:

$$L_i = L_{i-1} + L_{i-2} \text{ for } i > 1 \qquad (2)$$
$$L_0 = 2$$
$$L_1 = 1$$
$$L_2 = 3$$

There are three formulae relating the Fibonacci and Lucas numbers,

$$L_i = F_{i-1} + F_{i+1} \quad \text{for all integers i} \qquad (3)$$

and

$$5 F_i = L_{i-1} + L_{i+1} \text{ for all integers i} \qquad (4)$$

The third one is called Lucas factors of Fibonacci numbers,

$$F_{2i} = F_i \times L_i \qquad (5)$$

Therefore, the Lucas numbers are very closely related to the properties of Fibonacci numbers.

**3.3 The Lucas relationship within the canonical amino acids**

Previous studies show that it is possible to subdivide the set of 20 amino acids in many contrasting and overlapping ways. Organic synthetically, 20 canonical amino acids structurally and chemical compositionally vary from one to another by their different side-chain groups, which carry different number of carbon atoms. In the previous section, we discussed that, on the 28-gonal symmetric model of the genetic code system, a ratio of 15 [(P + S + F/L) or (T + M/I + N/K)] to 9 [(R + E/D) or (L + S/R)] equals to 1.6666, which is close to the Golden mean 1.6180339…, see Figure 5b.

In view of the structural chemistry of a biomolecule, the number of carbon atoms within an organic molecule or within a particular functional group of the molecule can often provide a clue to its biochemical origin. From a bio-organic synthesis point of view, 20 amino acids vary from one to another by their distinctive side-chain groups which carry a series of carbon atoms. The numbers of side-chain carbon atoms in canonical amino acids are composed of very simple numbers 0, 1, 2, 3, 4, 7 and 9, *i.e.*, ranging from 0 for glycine (G) to 9 for tryptophan (W) (Figure 6). A very interesting phenomenon is that any bigger number of the side-chain carbon-atoms (>2) in a bigger amino acid can be the sum of the two smaller numbers of the side-chain carbon-atoms in two smaller amino acids from anscester codons (eqn. 6). These two properties of the canonical amino acids are reminiscent of the Fibonacci-Lucas relationship in mathematics, see eqn's. 1, 2, and 6.

$$aa_i + aa_{i+1} = aa_{i+2} \qquad (6)$$

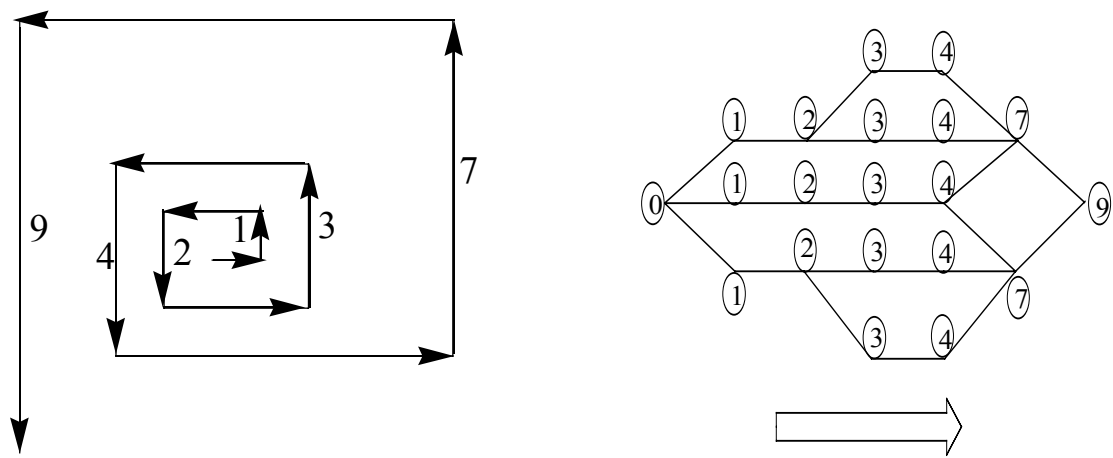

**Figure 6**.  Spiral rectangular relation of the side-chain carbon numbers within the 20 amino acid.

**Table 2.** A comparison of some Fibonacci numbers and Lucas numbers with side-chain carbon-atom numbers of the canonical amino acids.[a]

| *i* | 0 | 1 | 2 | 3 | 4 | 5 | 6 |
|---|---|---|---|---|---|---|---|
| Fibonacci numbers *or* $F_i$ | 0 | 1 | 1 | 2 | 3 | 5 | 8 |
| Lucas numbers or $L_i$ | 2 | 1 | 3 | 4 | 7 | 11 | 18 |
| *SCCAN* | 0 | 1 | 2 | 3 | 4 | 7 | 9 |
| *Type I amino acids* |  | C(1) |  | **V**(3); M(3) **E**(3); Q(3) | **L**(4); R(4) **I**(4); K*(4) | Y(7) | W(9) |
| *Type II amino acids* | **G**(0) | S(1) **A**(1) | T(2); **D**(2) N(2) | P(3) | H(4); K*(4) | F(7) |  |
| *Multiplicity* | 1 | 3 | 3 | 5 | 5 | 2 | 1 |

[a] *Type I amino acids* are amino acids recognized by class I aminoacyl-tRNA synthetases; *Type II amino acids* are recognized by class II aminoacyl-tRNA synthetases; The special case is lysine, which in some organisms is charged by a class I enzyme. Amino acids in bold are produced from Miller's prebiotic simulation experiment (Miller, 1987).

Three observations can be obtained from a comparison of the Fibonacci-Lucas numbers with the side-chain carbon-atom numbers within canonical amino acids in Table 2.  First, except Glycine (G), which possesses no side-chain carbon atom, and tryptophan (W), which carries 9 carbon atoms on its side-chain, 18 other canonical amino acids carry 1, 2, 3, 4, to 7 side-chain carbon-atoms, respectively, which exactly meet the initial 5 consecutive Lucas numbers, 2, 1, 3, 4 and 7.  Second, within the type I amino acids recognized by class I AARS's, the numbers of side-chain carbon atoms include 1, 3, 4 and 7, except W, following the Lucas rule, independent of the type II amino acids.  Within the type II amino acids recognized by class II AARS's, except G, the numbers of side-chain carbon atoms are 1, 2, 3, 4 and 7, being also consist with the Lucas sequence, independent of the type I amino acids.  Third, the above two observations merge to a suggestion that the majority of the 20 canonical amino acids selected in the genetic code are naturally selected by a double-Golden mean. Moreover, the sum of 1, 2, 3, 4 and 7 equals to 17, somehow in conformity with the number "17", which has been repetitively found in the amino-acid side-chain carbon-atom balance on the 28-gon model, as is pointed out in *Section 2.2*.

Seemingly, the above correlation between the amino-acid side-chain carbon-atom numbers (SCCAN) and Fibonacci-Lucas series prompts enquiry as to whether this quantitative property of the canonical amino acids appears to be a coincidence, especially considered together with the Miller's prebiotic simulation experiment (Miller, 1987), in which numerous small bio-molecules other than the standard amino acids are produced. However, in addition to the symmetric feature manifested by amino-acid side-chain carbon-atom number balance which is directed by the two classes of AARS's, very interestingly, the numbers "5" and "6", which are not in the Lucas series, are not selected into the amino-acid SCCAN category either.

Within the last decade, significant strides have been made toward understanding the molecular basis of the genetic code. One very important advance in this area was the definition of an amino acid property called the polar requirement (Woese, 1966). Recently, much progress has also been made in elucidating how specific interactions between amino acids and nucleic acids may have played an important role in the origin of the genetic code (Yarus, 1998; Szathmary, 1999). Whereas the evolutionary dynamic that shaped the code remains an enigma, and what property or properties of the amino acids the code actually reflects has long remained a mystery. In the present work, a pronounced similarity exists between the Lucas series (a numerical basis of the Golden mean) and amino-acid side-chain carbon-atom numbers, presumably suggesting that the amino acids selected into the genetic code system seems to follow the natural Golden mean by the Lucas series, when the 20 amino acids in the genetic code system are viewed as a whole.

## 4  Conclusion

Based on a detailed analysis of amino-acid side-chain carbon-atom numbers, it can be identified that the 28-gonal and rotational symmetric features within the internal relation of the 20 amino acids in the genetic code is stepwise intertwined with two classes of aminoacyl-tRNA synthetases. The two previously suggested evolutionary axes within the genetic code cooperatively overlap the three two-fold symmetry axes within two classes of aminoacyl-tRNA synthetases on the 28-gon model. Importantly, it has been revealed from this work that the natural selection of the side-chain carbon-atom numbers (1, 3, 4 and 7) in amino-acids is in an excellent agreement with a mathematical relationship, the Lucas series. These findings may provide new insight into biological understanding what principle governs the selection of the 20 canonical amino acids in the genetic code, and allow new opportunities in further theoretically exploring the information logic in the genetic code system.

**Abbreviations**

aa, amino acid; 20 amino acids are represented by A(Ala), P(Pro), V(Val), G(Gly), T(Thr), S(Ser), L(Leu), R(Arg), D(Asp), E(Glu), M(Met), I(Ile), F(Phe), C(Cys), W(Trp), H(His), Q(Gln), N(Asn), K(Lys) and Y(Tyr). "o" denotes "stop" codons.
SCCAN, Side-chain carbon-atom number; AARS, aminoacyl-tRNA synthetase.

(CMY, 2003/7/18)